# Topologically Charged Nodal Surface

Meng Xiao[*] and Shanhui Fan[+]

[1]*Department of Electrical Engineering, and Ginzton Laboratory, Stanford University, Stanford, California 94305, USA*

Corresponding E-mail: [*] mengxiao@stanford.edu and [+] shanhui@stanford.edu

Abstract: We report the existence of topologically charged nodal surface – a band degeneracy on a two-dimensional surface in momentum space that is topologically charged. We develop a Hamiltonian for such charged nodal surface, and show that such a Hamiltonian can be implemented in a tight-binding model as well as in an acoustic meta-material. We also identify a topological phase transition, through which the charges of the nodal surface changes by absorbing or emitting an integer number of Weyl points. Our result indicates that in the band theory, topologically charged objects are not restrict to zero dimension as in a Weyl point, and thus pointing to previously unexplored opportunities for the design of topological materials.

Classified by the dimension of band degeneracy, there are three classes of three-dimensional topological semimetals: Weyl and Dirac semimetals [1-17], nodal line semimetals[18-20], and nodal surface semimetals[21,22], where in the wavevector space the degeneracy occurs at a point, line, and surface respectively. However, while a Weyl point is topologically charged in that it carries a non-zero quantized total Berry flux, nodal lines and surfaces are not topologically charged in all previous works. Here we introduce an effective Hamiltonian for a topologically charged nodal surface, and show that such a Hamiltonian can be implemented in a tight-binding model and a realistic acoustic metamaterial. We also demonstrate a topological phase transition, where a nodal surface changes its charge by emitting or absorbing an integer number of Weyl points as the system parameters vary. Our work indicates that topologically charged objects in a band structure is not restricted to a point, but can have a much wider set of geometries, which points to additional possibility in topological material design.

The topology of both two-dimensional and three-dimensional band structures can be characterized by the Chern number. In a two-dimensional spinless system, nonzero Chern number is realized by breaking time reversal symmetry. For a three-dimensional system, the Chern number of a closed surface can be nonzero even without time reversal breaking. In this case, the topological charges, which result in nonzero Chern numbers, must come from band degeneracies. Weyl points [23,24] are band degeneracy points which are commonly known as the topological charges in the momentum space. A simple Hamiltonian of a Weyl point with charge $+1$ is

$$\hat{H} = q_x\sigma_x + q_y\sigma_y + q_z\sigma_z, \qquad (1)$$

where $q_x$, $q_y$ and $q_z$ are respectively the wavevector components along the *x*, *y* and *z* directions originating from the Weyl point, and $\sigma_x$, $\sigma_y$, and $\sigma_z$ are the Pauli matrices. The Weyl Hamiltonian in Eq. (1) is isotropic along all the directions in the momentum space, and hence the Berry flux coming out from this Weyl point is also isotropic. The total Berry flux coming out is $2\pi$, and hence the magnitude of Berry flux density decays as $1/(q_x^2+q_y^2+q_z^2)$ when away from the Weyl point as shown schematically in Fig. 1(a), where the red sphere represents the Weyl point and the arrows represent the direction and magnitude of the Berry flux density.

However, Weyl points are not the only kind of geometric objects in the momentum space that carries topological charge. Here we introduce a nodal surface that also carries nonzero topological charges. A Hamiltonian of a charged nodal surface with charge +1 is given by

$$\hat{H} = q_z\left(q_x\sigma_x + q_y\sigma_y\right) + q_z\sigma_z. \tag{2}$$

The eigenvalues and eigenvectors of this Hamiltonian are $E_\pm = \pm q_z\sqrt{1+q_\rho^2}$ and $v_\pm = \left(e^{-i\theta}\left(1\pm\sqrt{1+q_\rho^2}\right)/q_\rho, 1\right)$, respectively, where we have used $q_x = q_\rho\cos\theta$ and $q_y = q_\rho\sin\theta$, and the subscripts "+" and "−" correspond to the upper and lower bands, respectively. The two eigenvalues are degenerate at $q_z = 0$ and increase or decrease linearly with $q_z$ when away from the nodal surface, which indicates that the Hamiltonian in Eq. (2) represents a nodal surface. [21,22] The Berry connection is given by

$$\mathbf{A}_\pm = \frac{2+q_\rho^2 \pm 2\sqrt{1+q_\rho^2}}{q_\rho\left(2+2q_\rho^2 \pm 2\sqrt{1+q_\rho^2}\right)}\hat{\theta}, \tag{3}$$

which has only one non-vanishing component along the azimuthal direction. Hence the Berry curvature also possesses only one non-vanishing component which is along the *z* direction. If we consider two planes sandwiching this nodal surface, then the total Berry flux passing through each plane is $\pi$ as $\lim_{q_\rho\to\infty}\int q_\rho\mathbf{A}_\pm d\hat{\theta} = \pi$. Hence the Hamiltonian in Eq. (2) describes a nodal surface with topological charge +1. The nodal surface (transparent red surface) together with the Berry flux density (arrows) are shown in Fig. 1(b). The Berry flux density distribution is quite different from that of the Weyl point in Fig. 1(a), even though they both possess the same topological charge. Nodal surfaces can also possess other topological charges, as discussed in detail in Supplemental Material Sec. I. For a nodal surface with a higher topological charge, the Berry flux can be along other directions beside the *z* direction.

The Hamiltonian of Eq. (2) can be realized in a simple tight-binding model. The realizations of nodal surfaces have been discussed in Graphene networks [21] and a quasi-one-dimensional crystal family with nonsymmorphic lattice symmetries [22]. None of these works, however, showed a nodal surface that carries a non-zero topological charge. Here we consider a lattice system having a two-fold screw rotational symmetry along the $z$ direction $\tilde{C}_{2z}$, and time reversal symmetry $\mathcal{T}$. We define a compound anti-unitary symmetry operator $G_{2z} \equiv \mathcal{T}\tilde{C}_{2z}$, which, in real space, acts as

$$G_{2z}:(x,y,z,t) \mapsto (-x,-y,z+h/2,-t), \qquad (4)$$

where $h$ represents the unit cell size along the $z$ direction. It is easy to check that $G_{2z}^2 = -1$ for a Bloch wavefunction on the $k_z = \pi/h$ plane with arbitrary $k_x$ and $k_y$, where $k_x$, $k_y$ and $k_z$ are the wavevectors along the $x$, $y$ and $z$ directions, respectively. Therefore, $k_z = \pi/h$ forms a nodal surface due to a Kramers degeneracy on this surface.[22] An alternative compound symmetry $\Theta \equiv G_{2z}m_z$ also ensures the presence of a nodal surface, where $m_z$ is the mirror symmetry with respect to the $z$ direction.

The nodal surface as protected by either $G_{2z}$ or $\Theta$ symmetry, in general, may not have non-zero topological charge. In order to construct a charged nodal surface, one has to break either the inversion symmetry or time reversal symmetry or both as required in order to achieve a non-zero Berry flux. [25]. Therefore, the idea is to construct a charged nodal surface is to consider a system that has $G_{2z}$ or $\Theta$ symmetry, but breaks inversion or time-reversal symmetry.

The simplest model exhibits the symmetry of $G_{2z}$ or $\Theta$ consists of two sublattices. In Fig. 2(a), we consider such a tight binding model, which exhibits the $\tilde{C}_{2z}$ symmetry. Here red and blue spheres represent different sublattices, and their projections onto the $xy$ plane form a hexagonal lattice with a lattice constant $a$, with the red and blue spheres projected to the A and B sites of the honeycomb lattice. The lattice constant along the $z$ direction is $h$, and two sublattices are on the $z=0$ and $z=h/2$ planes. There is a coupling between two red (blue) spheres on neighbor A (B) sites at different z-planes with coupling strength $t_c$, as well as a coupling between nearest-neighbor red and blue sphere with strength $t_0$. The coupling strengths are assumed to be real and hence the system also exhibits time reversal symmetry. Thus the system has $G_{2z}$ symmetry. On the other hand, the system does not have inversion symmetry. The Hamiltonian of this tight binding Hamiltonian in the reciprocal space is given by:

$$\hat{H} = \begin{pmatrix} f(k_x, k_y, k_z) & g(k_x, k_y, k_z) \\ [g(k_x, k_y, k_z)]^* & f(k_x, k_y, -k_z) \end{pmatrix}, \quad (5)$$

with

$$f(k_x, k_y, k_z) = 2t_c \left[ \cos(k_x a + k_z h) + 2\cos\left(\frac{k_x a}{2} - k_z h\right) \cos\left(\frac{\sqrt{3}k_y a}{2}\right) \right], \quad (6)$$

$$g(k_x, k_y, k_z) = 2t_0 \cos\left(\frac{k_z h}{2}\right) \left[ 2\cos\left(\frac{k_x a}{2}\right) \exp\left(i\frac{\sqrt{3}k_y a}{6}\right) + \exp\left(-i\frac{\sqrt{3}k_y a}{3}\right) \right]. \quad (7)$$

It is easy to show that near $k_z = \pi/h$, both $f(k_x, k_y, k_z) - f(k_x, k_y, -k_z)$ and $g(k_x, k_y, k_z)$ are proportional to $k_z - \pi/h$, which proves that such a tight-binding model possesses a nodal surface at $k_z = \pi/h$, as required by the $G_{2z}$ symmetry as discussed above. Figure. 2(b) shows the band structure along several directions in the reciprocal space, which shows clearly the features of a nodal surface: The bands are degenerate at $k_z = \pi/h$ for arbitrary $k_x$ and $k_y$, and the dispersion is linear when away from the nodal surface. Besides the presence of a nodal surface, we also note that the band dispersions are linear along all the directions at the *K* point, indicating the existence of a Weyl point at the *K* point.

Keeping to the lowest order, the Hamiltonian in Eq. (5) near $H \equiv (4\pi/3a, 0, \pi/h)$ and $H' \equiv (-4\pi/3a, 0, \pi/h)$ can be written as

$$\hat{H} = 3t_c \mathbf{I} + \frac{\sqrt{3}}{2} \left[ t_0 a h q_z \left( \pm q_x \sigma_x - q_y \sigma_y \right) \mp 6 t_c h q_z \sigma_z \right], \quad (8)$$

where $\mathbf{I}$ is the $2 \times 2$ identity matrix, and the upper and lower signs represent the Hamiltonian near the *H* and *H'* points, respectively. Eq. (8) and Eq. (2) share the same form, and hence they should both possess the same charge (for both *H* and *H'* points). Therefore, the nodal surface at $k_z = \pi/h$ of this tight-binding model is topologically charged. This conclusion is also consistent with the presence of Weyl points at *K* and *K'*. The charges of these two Weyl points are the same. Since the total charges inside the Brillouin zone must vanish, the charges of Weyl points must be compensated by the charges of the nodal surfaces, as these points and surfaces are the only band-degenerate features in the Brillouin zone.

As a direct numerical check that the nodal surface is indeed charged, Fig. 2(c) shows the Chern number as a function of $k_z$ for different $k_z$ planes of the upper band (red) and the lower band (blue). The change of Chern number at $k_z = 0$ is due to the presence of two Weyl points while the change of Chern number at $k_z = \pi/h$ is due to the charged nodal surface. Hence we can conclude that this nodal surface possesses topological charge +2. As the Chern number for a fixed $k_z$ can be nonzero, there exist one-way edge states when the system is finite according to the bulk-edge correspondence. Conversely, the existence of a one-way edge state at a fixed $k_z$ for a finite system is a direct indication that the Chern number at such a $k_z$ is non-zero. [26] In Fig. 2(d), we consider a strip geometry which is periodic along the $x$ and $z$ directions and finite along the $y$ direction with a zigzag boundary. We fix $k_z h = \pi/2$ and show the band structure as a function of $k_x$. The cyan area represents the projected bulk band. The red and blue curves represent the one-way edge states localized on the upper and lower boundaries, respectively. The one-way nature of these edge states can also be seen with a time-dependent simulation of the tight-binding model in a finite system. (Supplemental Material Sec. II) The existence of the one-way edge states confirms the nontrivial topology of this nodal surface.

In general, in a system with a $G_{2z}$ symmetry, there is always a nodal surface at $k_z h = \pi$. This nodal surface may be either topologically trivial or non-trivial, depending on system parameters. Here we show that the transition between a topologically trivial and non-trivial nodal surfaces involves the "absorption" and "emission" of Weyl points by the nodal surface. As an illustration, we add an additional onsite interlayer coupling term $3\sqrt{3}\xi t_c \sin(2k_z h)\sigma_z/2$ to the Hamiltonian in Eq. (5). This term can be realized with onsite hopping between next nearest layers with opposite hopping phases at different sublattices. [27] Since this term vanishes at $k_z = 0$ and $k_z = \pi/h$, it doesn't lift the degeneracy of the nodal surface and it also preserves the existence of Weyl point at $K$ and $K'$. Here $\xi$ characterizes the hopping strength and the numerical factors in the term is chosen to ensure that the topological transition occurs at $\chi = \pm 1$ (See Supplemental Material Sec. I) The topological charge distribution for $\xi < 1$ is shown in Fig. 3(a), where the red transparent plane and blue spheres represent the positively charged nodal surface and Weyl points with charge $-1$, respectively. At $\xi = 1$, the $q_z \sigma_z$ term in Eq. (8) vanishes at $H$ and $K'$. However, the topological charges in fact remains unchanged since to the next order there is still the $q_z^3 \sigma_z$ term. (See Supplementarl Material Sec. I) At $\xi > 1$, the nodal surface becomes topologically trivial with two charge $+1$ Weyl points emerging and moving away from it. The Weyl point at $K'$ decomposes into three Weyl points: two with charge $-1$ go away and one with charge $+1$ stays. This charge distribution is shown schematically in Fig. 3(b), where we use the gray transparent surface to represent the topologically trivial nodal surface. A movie shows this topological transition process can be seen in Supplemental Material Movie S. 1. For the Hamiltonian shown here, the only geometric object where degeneracy can occur are the Weyl points and the nodal surfaces. The process through which the nodal surface changes its charge, as illustrated above, is quite general for this class of Hamiltonian– For a process through which the nodal surface is preserved, the nodal surface can only change its charge by emitting or absorbing Weyl points.

We now proceed to show that topologically nontrivial nodal surfaces can be found in acoustic metamaterials with full wave simulations using COMSOL[28]. We consider systems consisting of acoustic resonance cavities with connection tubes controlling the hopping strength[9]. This acoustic system has been previously used experimentally to demonstrate various topological concepts.[29-31] Figure 4(a) shows the unit cell used in the simulation, where blue and yellow represent the surfaces where sound hard boundary condition and periodic boundary condition are applied, respectively. The system is filled with air. It is easy to see that this unit cell possesses the $G_{2z}$ symmetry. Figure 4(b) shows the corresponding band structure. All bands are two-fold degenerate at the surface of $k_z = \pi/h$ and exhibit linear dispersion away near the surface. The $k_z = \pi/h$ surface is therefore a nodal surface. The existence of a Weyl point at the $K$ point is also confirmed as the band dispersion are linear along all the high symmetry direction when away near the $K$ point. In order to demonstrate that this nodal surface is topologically nontrivial, we focus on the two red bands in Fig. 4(b) which possesses a relatively larger band gap when $k_z \neq \pi/h$ or $0$, and then consider a strip of unit cell as shown in Fig. 4(c), where the left panel shows the sketch of the strip and the right panel shows the details of the boundary. The strip is periodic along the *x* and *z* directions, and confined with sound hard boundaries along the *y* direction. For a fixed $k_z = p/2h$, the band structure as a function of $k_x$ is shown in Fig. 4(d). Cyan area shows the projection of the bulk band, and red and blue curves represent the dispersion of the surface states localized on the upper and lower edges, respectively. As there are no other band degenerate points except for the Weyl points and the nodal surface, the existence of one-way edge states indicates that the nodal surface is topologically nontrivial.

Our work here indicates the existence of a charged nodal surface, a new type of geometric object in momentum space that carries topological charge. System possessing such a charged nodal surface represents a new class of topological semimetal. For a nodal surface with zero charge, lifting the $G_{2z}$ symmetry results in the creation of a band gap. In contrast, for the system considered here that has a topologically charged nodal surface, when the symmetry $G_{2z}$ is broken, charged nodal surface will become Weyl points and the semimetal property is still topologically protected. We have provided a physical implementation of an acoustic metamaterial that possesses such a topologically charged nodal surface. We believe that such a topologically charged nodal surface can be realized in electronic and electromagnetic systems as well.

This work is supported by the U. S. Air Force of Scientific Research (Grant No. FA9550-12-1-0471), and the U. S. National Science Foundation (Grant No. CBET-1641069).

# Figures

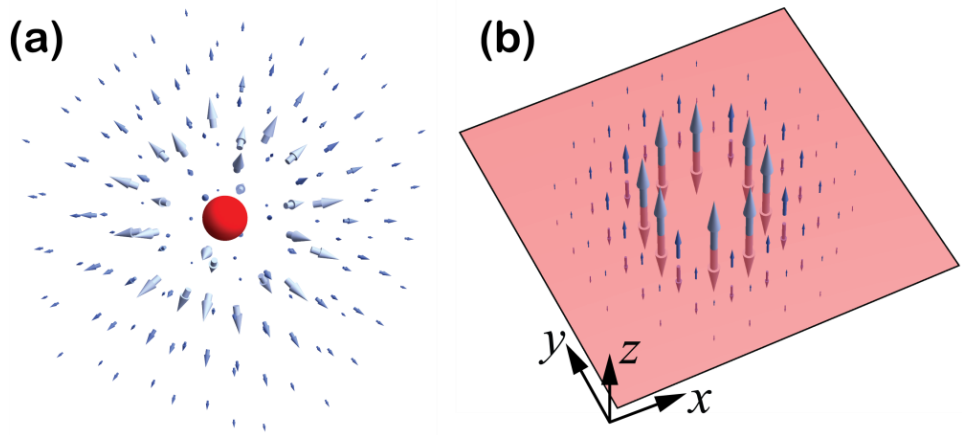

Fig. 1 (color online) Berry flux density distributions of a Weyl point (a) and a charged nodal surface (b). The red sphere and transparent red plane represent the Weyl point and charged nodal surface, respectively. The arrows represent the direction and amplitude of the Berry flux density.

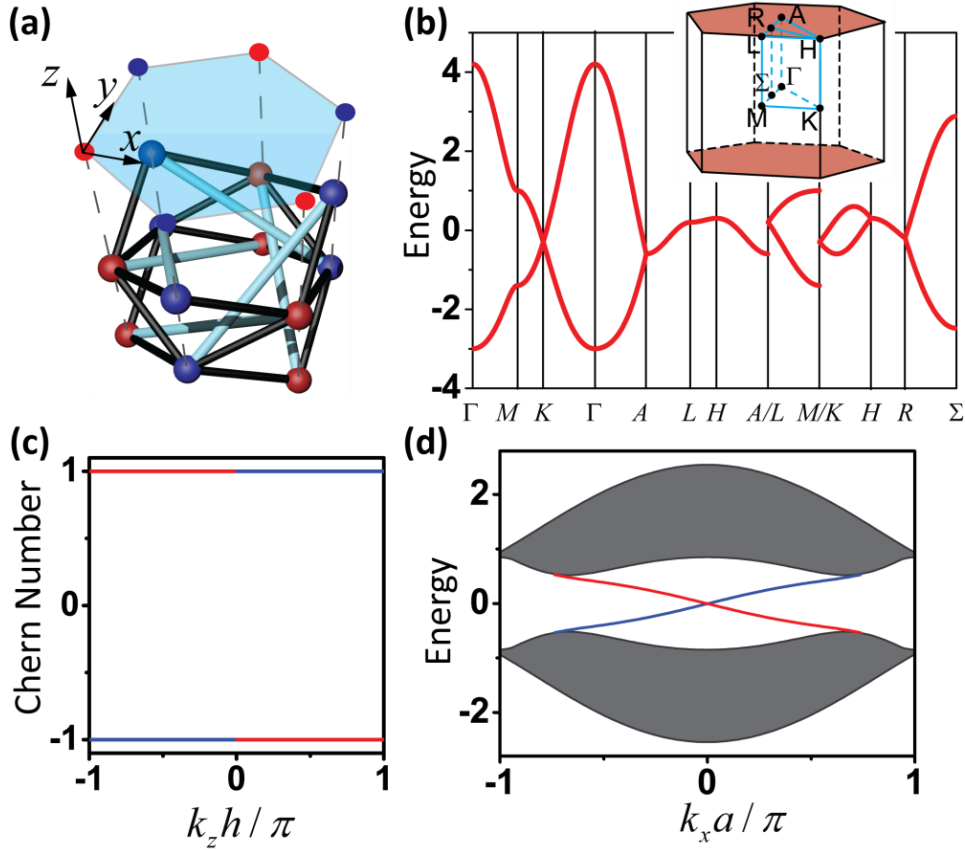

Fig. 2 (color online) (a). A sketch of the tight binding model. Each unit cell consists of two lattice sites, represented by red and blue spheres respectively. Their projections onto the *xy* plane form a hexagonal lattice with a lattice constant *a*. The lattice constant along the *z* direction is *h*, and the red and blue spheres are located on $z=nh$ and $z=(n+1/2)h$ plane, respectively, where *n* is an integer. Bonds represent hopping between different sites. Hopping strengths of the cyan and black bonds are given by $t_c$ and $t_0$, respectively. This model exhibits the $\tilde{C}_{2z}$ symmetry. (b). The band structure of the tight binding model in (a), where $t_c = 0.1$ and $t_0 = 0.6$. The reciprocal space is shown in the inset with positions of relevant high symmetry points marked. Here the red transparent surface represents the nodal surface on which two bands are always degenerate. (c). The Chern numbers as a function of $k_z$. Red and blue curves represent the Chern numbers for the upper and lower bands, respectively. (d). Projected band (grey) for a finite strip of the periodic system shown in (a). The strip is periodic along the *x* and *z* directions, and is terminated with a zigzag boundary on both ends along the *y* direction. We fix $k_z h = \pi/2$. Blue and red curves represent the surface states localized at the lower and upper zigzag boundaries, respectively.

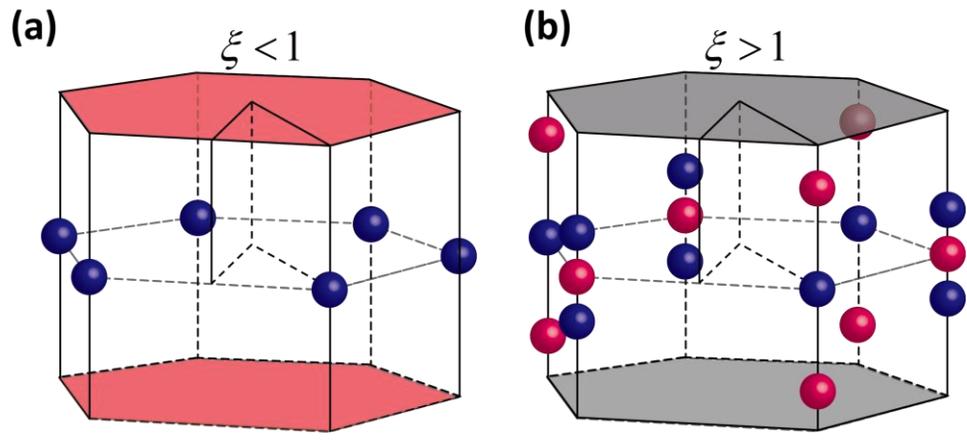

Fig. 3 (color online) Topological charge distributions before the topological transition (**a**) and after the topological transition (b). Here red and blue spheres represent Weyl points with positive and negative charges, respectively. Red and gray surfaces represent the nodal surfaces with positive and zero charges, respectively.

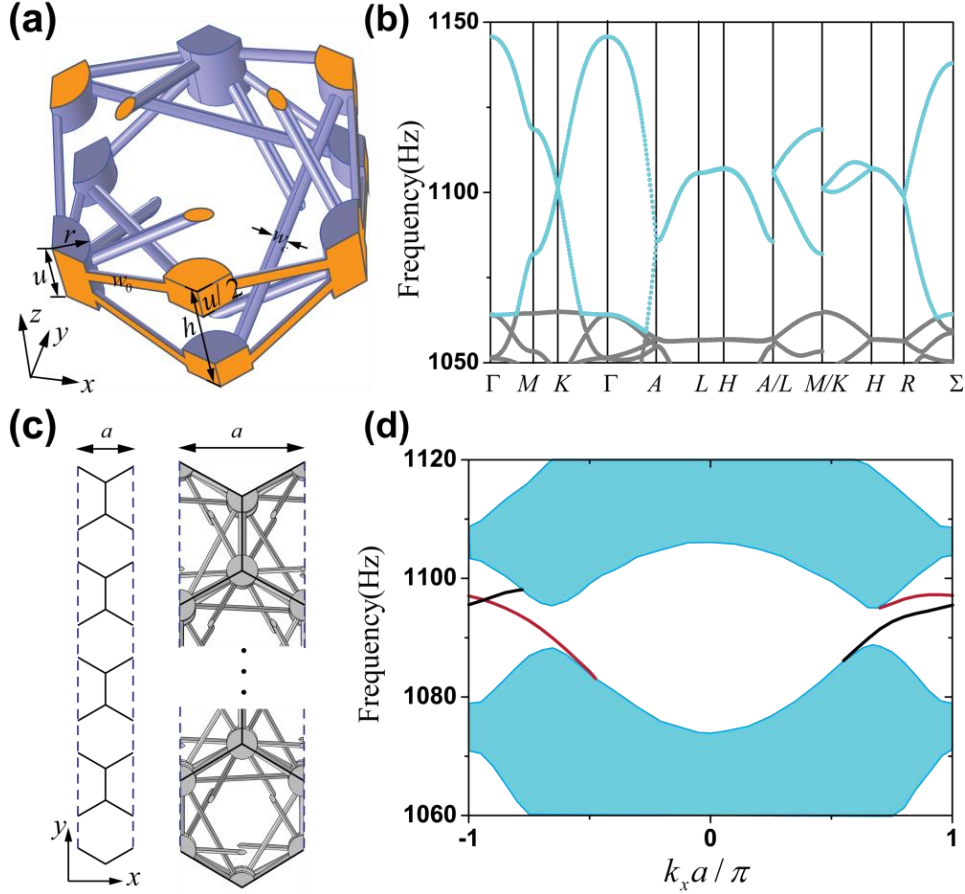

Fig. 4 (color online) (a). A unit cell of the acoustic system under consideration, where blue and yellow represent the surfaces where hard boundary and periodic boundary conditions are applied, respectively. The system is filled with air (density 1.29 $kg/m^3$ and speed of sound 343m/s ) (b). The band structure along various directions in the first Brillouin zone, where cyan corresponds to the projected bulk bands studied in (d). (c). The unit cell used to calculate the projected band in (d). The left panel shows the schematic setup and the right panel shows the details of the boundary. The strip is periodic along the $x$ and $z$ directions, and confined with the sound hard boundary condition along the $y$ direction. The number of unit cell along the $y$ direction is chosen to be large enough such that the dispersions of the surface states no longer change as the number of unit cell further increases. (d). The cyan area shows the projection of the bulk bands as highlighted in (b), and red and black curves represent the dispersion of the surface states localized on the upper and lower edges, respectively. The parameters used are $a=40 cm$, $h=24 cm$, $r=5 cm$, $w_0=2.4 cm$, $w_c=1.6 cm$, and $u=12 cm$. We fix $k_z h=\pi/2$ in d.